\documentclass[onecolumn]{aastex63}

\shorttitle{Parsec-scale properties of TXS~0506+056}
\shortauthors{Li et al.}

\usepackage{amsmath,url}
\usepackage{hyperref}

\begin{document}

\title{The parsec-scale jet of the neutrino-emitting blazar TXS~0506+056}

\author{Xiaofeng Li} \affil{Shanghai Astronomical Observatory, Key Laboratory of Radio Astronomy, Chinese Academy of Sciences, Shanghai 200030, China} \affil{Shanghai Tech University, 100 Haike Road, Pudong, Shanghai, 201210, China} \affil{University of Chinese Academy of Sciences, 19 A Yuquan Road, Shijingshan District, Beijing 100049, China}

\author{Tao An} \affil{Shanghai Astronomical Observatory, Key Laboratory of Radio Astronomy, Chinese Academy of Sciences, Shanghai 200030, China} 

\author{Prashanth Mohan} \affil{Shanghai Astronomical Observatory, Key Laboratory of Radio Astronomy, Chinese Academy of Sciences, Shanghai 200030, China}

\author{Marcello Giroletti} \affil{Istituto Nazionale di Astrofisica, Istituto di Radioastronomia, via Gobetti 101, I40129, Bologna, Italia}

\correspondingauthor{Tao An}
\email{antao@shao.ac.cn}

\def\deg{$^\circ$}
\def\fermi{{\it Fermi}}
\def\txs{TXS\, 0506+056} 
\def\pflux{\mathrm{ph}\,\mathrm{cm}^{-2}\,\mathrm{s}^{-1}}

\begin{abstract}
Recently the IceCube collaboration detected very high energy (VHE) neutrinos and associated them with the blazar \txs{}, 
raising a possible association of VHE neutrinos with this and other individual blazars. Very Long Baseline Interferometry (VLBI) is so far the only technique enabling the imaging of the innermost jet at milli-arcsec resolution (parsec scale), where the high energy emission possibly originates from. Here, we report on the radio properties of the parsec scale jet in \txs{} derived from the analysis of multi-epoch multi-frequency archive VLBI data. 
The half opening angle of the jet beam is about 3.8\degr, and the jet inclination angle is about 20\degr. The overall jet structure shows a helical trajectory with a precessing period of 5--6 years, likely originating from instabilities operating at parsec scales.  
The calculated beaming parameters (Doppler boosting factor, bulk Lorentz factor) suggest a moderately relativistic jet. 
The pc-scale magnetic field strength is estimated in the contexts of core-shift and variability, and is in general agreement in the range of 0.2 - 0.7 G. And it is found to decrease from a relatively larger value during the quiescent period before the ongoing flare. This suggests a conversion of magnetic field energy density to particle energy density that help accelerate injected particles at the jet base and result in variable shocked emission. The neutrino event could be associated with the onset of energetic particle injection into the jet. This scenario then supports the lepto-hadronic origin of the VHE neutrinos and $\gamma$-ray emission owing to a co-spatial origin. 
\end{abstract}

\keywords{BL Lacertae objects: individual (TXS 0506+056) - cosmic rays - galaxies: jets - gamma rays: galaxies - neutrinos - radiation mechanisms: non-thermal}

\section{Introduction} \label{sec:1}

The blazar class of active galactic nuclei (AGN) are characterized by  highly beamed relativistic jets (collimated with half opening angle $<$ 5$^\circ$) directed very close to the observer's line of sight \cite[e.g.][]{1995PASP..107..803U,2017MNRAS.468.4992P}. Their Doppler beamed strongly variable emission spans a broad range of wavelengths from the radio to $\gamma$-rays \cite[e.g.][]{1996ApJ...463..444S,2012MNRAS.425.1357G}, and can be accompanied by energetic cosmic ray particles and neutrinos \cite[e.g.][]{1993A&A...269...67M,1995APh.....3..295M,1996SSRv...75..341S,2002RPPh...65.1025H,2019NatAs...3...88G}. The innermost parsec-scale region of these energetic AGN hosts the radio core and collimated jet composed of stationary and apparently superluminal compact unresolved jet components \cite[e.g.][]{1997ARA&A..35..607Z,2014ApJ...787..151C,2013AJ....146..120L, 2016AJ....152...12L}. These peculiar properties render high resolution studies of this region of prime importance in the contexts of jet launching and collimation \cite[e.g.][]{2015ApJ...798..134H}, emission processes and contributing components \cite[e.g.][]{2018MNRAS.473.3638G}, interaction with the local environment \cite[e.g.][]{2019ApJ...873...11J,2020NatCo..11..143A}, and relation to mechanisms enabling multi-wavelength variability \cite[e.g.][]{2018ApJ...854...17L}.

The broadband spectral energy distribution of blazars is dominated by non-thermal emission, exhibiting a double-peaked structure. The low-frequency peak from radio to soft X-ray bands originates from relativistic electrons, while the high-frequency component from X-ray to TeV $\gamma$-ray bands can originate from distinctive physical scenarios. In the leptonic scenario \cite[e.g.][]{2008MNRAS.385..283C,2013ApJ...768...54B}, the high energy emission originates from the synchrotron self-Compton \cite[up-scattering of low energy synchrotron photons by the same population of relativistic electrons, e.g.][]{1992ApJ...397L...5M} and/or from the inverse Compton up-scattering of external seed photons from the accretion flow, broad line region and torus \citep{1994ApJ...421..153S,2009MNRAS.397..985G,2016ApJ...830...94F}. In the hadronic scenario \cite[e.g.][]{2013ApJ...768...54B}, the high energy emission consists of synchrotron from decelerating protons and from the particle decay originating in hadronic interactions \citep[e.g.][]{1992A&A...253L..21M,2019ApJ...876..109Z}. In a hybrid lepto-hadronic scenario, the acceleration zone at the jet base hosts strong shocks which can accelerate electrons and protons, with the former producing the observed synchrotron emission and enabling the subsequent hadronic interactions and pair cascades involving the latter thus producing high energy emission \citep[e.g.][]{1993A&A...269...67M, 2001APh....15..121M}.

The blazar TXS 0506+056 \cite[$z = 0.3365$,][]{2018ApJ...854L..32P} became a target of intense multi-wavelength monitoring owing to the detection of a 290 TeV neutrino IceCube-170922A coincident with its direction and arrival time during a $\gamma$-ray flare \citep{2018Sci...361..147I,2018Sci...361.1378I}. \txs{} was also identified as the most plausible origin for an earlier, 2014/2015 neutrino flare during a jet dominated low-hard state \citep{2018MNRAS.480..192P}, providing evidence for the blazar jet acting as an accelerator of cosmic ray particles (which produce the neutrinos). This discovery opens up the possibility of an association between blazars and high energy neutrinos, especially during the flaring epochs as sites for the origin of cosmic neutrinos \cite[e.g.][]{2018ApJ...854...54R,2018ApJ...865..124M}, and necessitated studies on jet emission mechanisms that can produce high energy cosmic rays and neutrinos \cite[e.g.][]{2018ApJ...864...84K,2019MNRAS.483L..12C}. As neutrinos are uniquely produced in the hadronic and lepto-hadronic scenarios during the interactions and particle decays, their detection during or preceding $\gamma$-ray flares suggests a co-spatial origin and consistency with these emission scenarios.

Radio very long baseline interferometry (VLBI) observations with milli-arcsecond (mas) resolution of \txs{} and other putative neutrino associated blazars probe the pc-scale morphology and environment of the jets, and help address the association of the radio jet with the $\gamma$-ray emission and locate the production sites of neutrinos \cite[e.g.][]{2019A&A...630A.103B,2020A&A...633L...1R,2020arXiv200100930P}. The analysis of 15 GHz data from the very long baseline array (VLBA) spanning the years 2009 to 2018 \citep{2019A&A...630A.103B} identifies a pc-scale strongly curved Doppler beamed and precessing jet with the neutrino production site likely to be further downstream of the $\gamma$-ray emitting region; peculiar jet component kinematics and flux density evolution are interpreted as originating in either a single strongly curved jet or in a jet-jet interaction involving two jets supported by Lense-Thirring precession due to a misalignment between the angular momentum in the inner hot accretion flow and the black hole spin, such as that enabled by a supermassive black hole (SMBH) binary. Higher frequency 43 GHz VLBA observations following the neutrino event probe the millimeter-VLBI core \citep{2020A&A...633L...1R} closer to the jet base; the observations indicate a spine-sheath structured jet with the electrons and protons accelerated in the spine up-scattering photons from the slower sheath, thus suggesting a co-spatial origin. The analysis of single dish 15 GHz radio flux densities from the Owens Valley Radio Observatory (OVRO) spanning between 2008 to 2018 indicates that the core of \txs{} is in a highly flaring state coincident with the neutrino event 170922A \citep{2019MNRAS.483L..42K,2019A&A...630A.103B}; the same study also identifies a curved jet region at $<$ 10 pc contained in a Doppler beamed helical jet by analyzing the archival 15 GHz VLBA observations. These complex clues necessitate continued follow-up VLBI observations and studies of archival data to clarify the nature of the pc-scale jet and understand the origin of the high energy emission and neutrinos. 

The 15 GHz VLBA data investigated in \citet{2019MNRAS.483L..42K} and \citet{2019A&A...630A.103B} is re-processed here for the detailed investigation of the pc-scale source morphology, jet kinematics (proper motion of jet components, position angle, bulk Lorentz factor, inclination angle) and flux density variability which are of high relevance in the above contexts. Near simultaneous multi-band radio VLBI data at 8 GHz, 24 GHz and 43 GHz in addition to the 15 GHz data spanning mainly the years 2009 to 2018 (with one during 2003) are used to estimate the radio core distance and associated magnetic field strength at the jet base. This adds to the observational constraints available for interpretation of the complex jet physics compared to the above studies of this source. The data analysis is discussed in Section \ref{sec:2}. Section \ref{sec:3} discusses the jet morphology, kinematics, beaming properties, the parsec-scale magnetic field strength and flux density variability. The important results are then summarized in Section \ref{sec:5}. At a redshift of $z = 0.3365$, in the $\Lambda$ cold dark matter cosmological model with Hubble constant $H_0 = 70\, {\rm km}\, {\rm s}^{-1}\, {\rm Mpc}^{-1}$, matter density $\Omega_{\rm M} = 0.3$, and dark matter density $\Omega_{\Lambda} = 0.7$, an angular size of 1 mas corresponds to a projected physical size of 4.83 pc, and 1 mas yr$^{-1}$ corresponds to $21.0\, c$.

\section{Data reduction and analysis} \label{sec:2}

We obtained all available VLBI data from archival databases, which includes 26-session observations spanning from 2003 September 13 to 2018 December 16. The 8 GHz VLBI data is acquired from the VLBI Calibrator Survey (VCS) project\footnote{Astrogeo VLBI FITS image database maintained by Leonid Petrov: \url{http://astrogeo.org/}}, and, the 15 GHz data from the Monitoring Of Jets in Active galactic nuclei with the Very long baseline array (VLBA) Experiments \citep[MOJAVE; ][]{2009AJ....137.3718L} survey archive\footnote{\url{http://www.physics.purdue.edu/astro/MOJAVE/index.html}}. The MOJAVE data has been calibrated using the Astronomical Image Processing System (AIPS) software package \citep{2003ASSL..285..109G}, and the VCS data is processed in the {\sc PIMA} software package \citep{2011AJ....142...35P}. We only performed self-calibration for these data sets in the {\sc Difmap} software package \citep{1997ASPC..125...77S} to eliminate the residual phase {\b and amplitude} errors and to increase the dynamic range (the radio of the peak intensity to the image noise) of the images.

The 24 and 43 GHz data are from the astrometry project of high-frequency celestial reference frame \citep[CRF;][]{2010AJ....139.1695L}. We downloaded the raw data from the National Radio Astronomy Observatory (NRAO) data archive\footnote{\url{https://archive.nrao.edu/archive/advquery.jsp}} and calibrated the data with AIPS following standard procedures described in the AIPS cookbook. Further details on the data reduction and estimation of image parameters are presented in Appendix \ref{sec:A1}.
The 15-GHz images with an optimal combination of sensitivity and resolution are used for the jet morphology and kinematics analysis in Section \ref{sec:3.1} and \ref{sec:3.2}. The core parameters derived from simultaneous 24 and 43 GHz data, as well as the 8 and 15 GHz data in close epochs, are used for the core shift and magnetic field strength calculations in Section \ref{sec:Bfield}. Due to the different resolution and opacity effect at different frequencies, we do not attempt a cross-identification of the jet component models fit at different frequencies. 

\section{Results and discussion} \label{sec:3}

\subsection{Jet emission structure and morphology} \label{sec:3.1}

\begin{figure*}
\centerline{\includegraphics[width=1.0\linewidth]{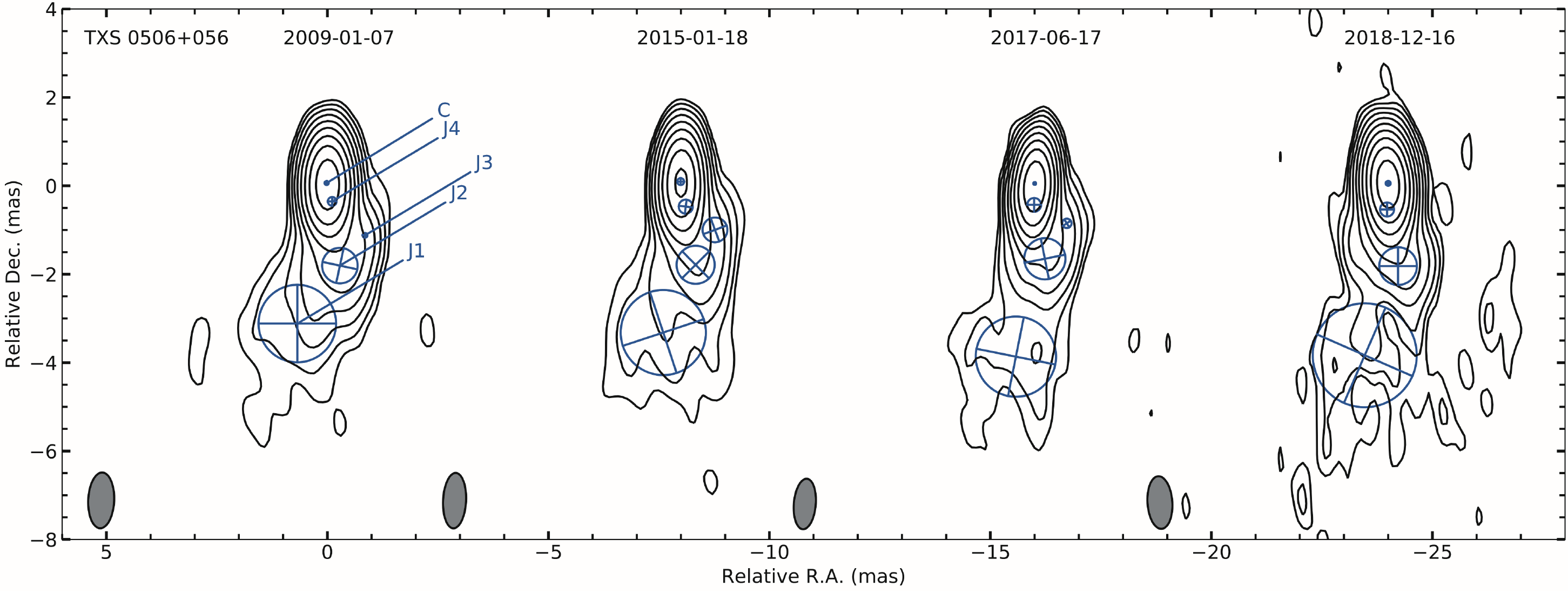}}
\caption{Parsec-scale radio structure of \txs{} at 15 GHz. 
The image parameters are presented in Table \ref{tab:obs}. 
The VLBI components are marked as blue-colored circles and with text labels.
A total of 17 epochs of VLBI data were processed for our analysis, among which four images are representative of the jet structure evolving with time. 
They are: 
2009 January 7 --- corresponding to the first epoch of the VLBI data; 
2015 January 18 --- in the mid of the time period in which $3.5\sigma$ evidence for neutrino emission was detected from the direction of \txs{} \citep{2018Sci...361..147I};
2017 June 17 --- two months before the 2017 September neutrino flare episode \citep{2018Sci...361.1378I};  and 
2018 December 16 --- the last epoch of the available VLBI data.
A prominent jet bending is detected around 2 mas from south to the southeast at the location of J2. 
A significant change of the jet structure is found between these epochs. 
Quantitative analysis of the jet proper motion and geometry is referred to Section \ref{sec:3}. 
The sensitivity of individual data sets are diverse. For a consistent comparison, a uniform value of 1 mJy beam$^{-1}$ is adopted for the first contour of each image. Other contours increase in steps of 2 mJy beam$^{-1}$. 
The grey-shaded ellipse to the bottom-left of each image represents the restoring beam.
}
\label{fig1}
\end{figure*}

\begin{figure}
\centerline{\includegraphics[width=\linewidth]{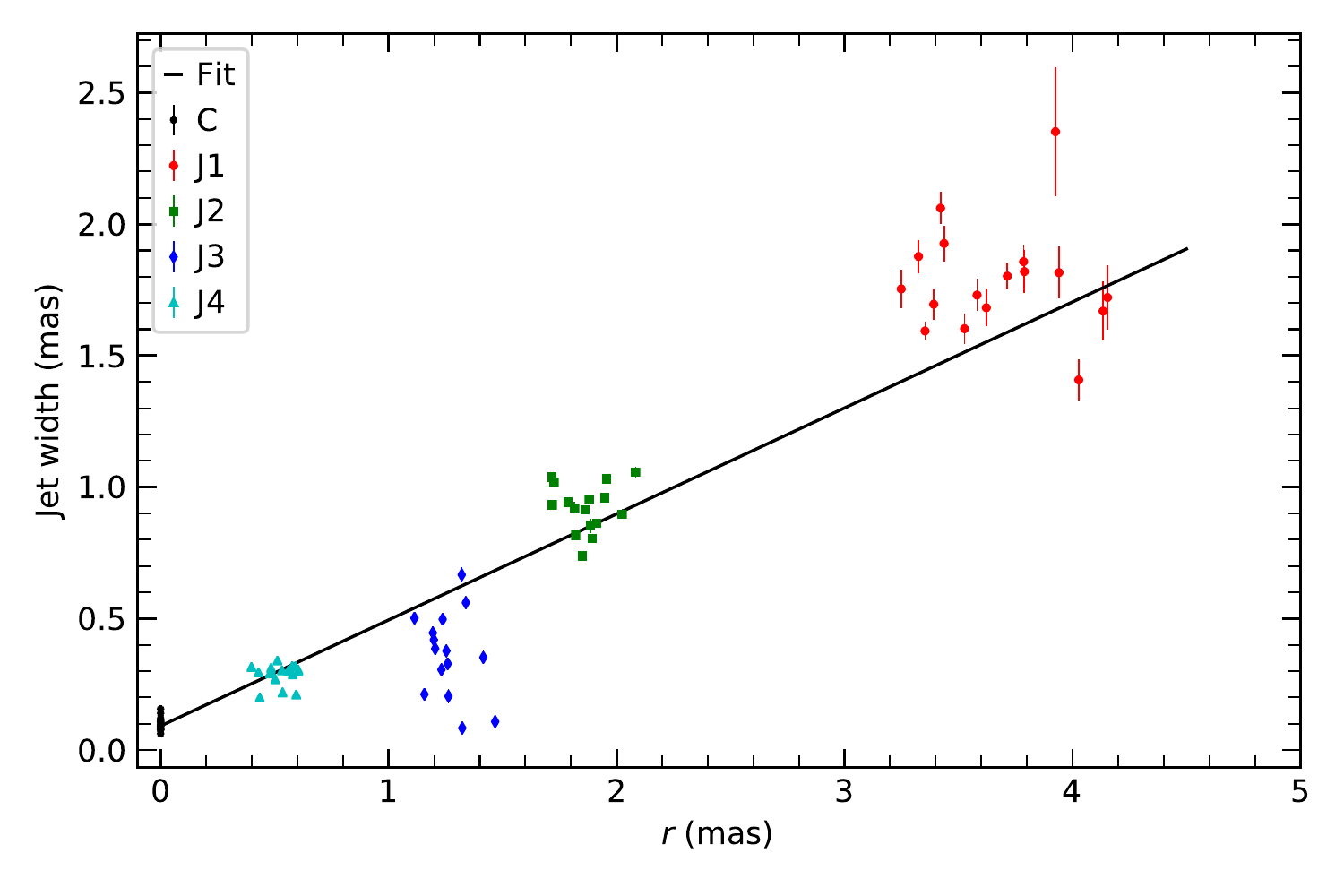}}
\caption{Opening angle of the jet body. The data points are derived from 15 GHz multi-epoch observations and presented in Table \ref{tab:comp}.}
\label{fig2}
\end{figure}

Figure \ref{fig1} shows the radio jet morphology of \txs{} on four representative epochs at 15 GHz. 
A core and four jet components (named as J1 to J4 from the outermost inwards) are fitted. This nomenclature is different from the previous VLBI studies \citep{2019MNRAS.483L..42K,2020A&A...633L...1R} and ensures a uniformity in the description of emitting jet components in this study. The largest detectable jet extent is related to the quality of individual images. The {\it rms} noise levels in most images are in the range of 0.1--0.3 mJy beam$^{-1}$, except the last two images which show slightly higher noise levels of 0.4 and 0.5 mJy beam$^{-1}$ (Table~\ref{tab:obs}).  In order for a proper comparison, we fixed the lowest contours in Figure~\ref{fig2} to a uniform value of 1 mJy beam$^{-1}$, which corresponds to 2--5 times the {\it rms} noise (Table \ref{tab:obs}). 
The maximal jet length at 15 GHz was observed on 2018 May 31, which we estimate to be about 4.2 mas in projection on the plane of the sky.

The jet shows a prominent bending morphology, as seen in Figure~\ref{fig1}. The jet body extends first to the south, then turns to southeast at about 1.5--2 mas away from the core. Two jet components J2 and J3 are detected at the section of jet bending. The 8 and 5 GHz data indicate a consistent curved jet structure as inferred from the 15 GHz images. The position angles of the innermost jet component J4 show a distinct gradual change from $\sim 197\degr$ (or $-163\degr$ in Table \ref{tab:comp}) in 2009 to $178\degr$ in 2018. Similar changes of position angles are also seen in J2 and J3. From Table \ref{tab:obs}, it can be inferred that the restoring beam of each individual image is similar due to the almost identical ($u$, $v$) coverage of these observations. The distinct change in the jet structure may not simply originate from the image quality systematics or deconvolution beams but may be genuinely related to the jet dynamical structure.

The sizes of the fitted jet components increase with the radial distance from the center. The outermost component J1 is the most extended with the largest size, and the innermost J4 has the smallest size. Therefore a conical geometry could be a good description of the \txs{} jet.
Since the jet body is curved, the opening angle of the jet may not be appropriately represented by the envelope containing all jet components.
Figure \ref{fig2} shows the jet width ($w_r$) versus radial distance of the jet component from the core ($r$) for all epoch data at 15 GHz. 
The jet width is estimated based on the jet component size from the model fitting of the visibility data with circular Gaussian models (refer to Table \ref{tab:comp}) rather than from the fitting of the brightness cross section perpendicular to the jet ridgeline in the image.
This is as the former can yield reduced errors due to sparse sampling in the $u-v$ space and an increase in the weights assigned to bright structures.
The scatter in the data points for each component, mainly along the jet width axis owes to errors arising from the model fitting of their corresponding brightness distributions, and random errors.
The jet width at increasing radial distances is well fitted with a linear function, supporting the inference of a conical jet body. 
The projected half opening angle of the jet beam $\tan2\phi=\Delta{}w_r/\Delta{}r$ from which $\phi = 11.0\pm0.3\degr$. 

\subsection{Jet component kinematics} \label{sec:3.2}

\begin{figure}
\centerline{\includegraphics[width=\linewidth]{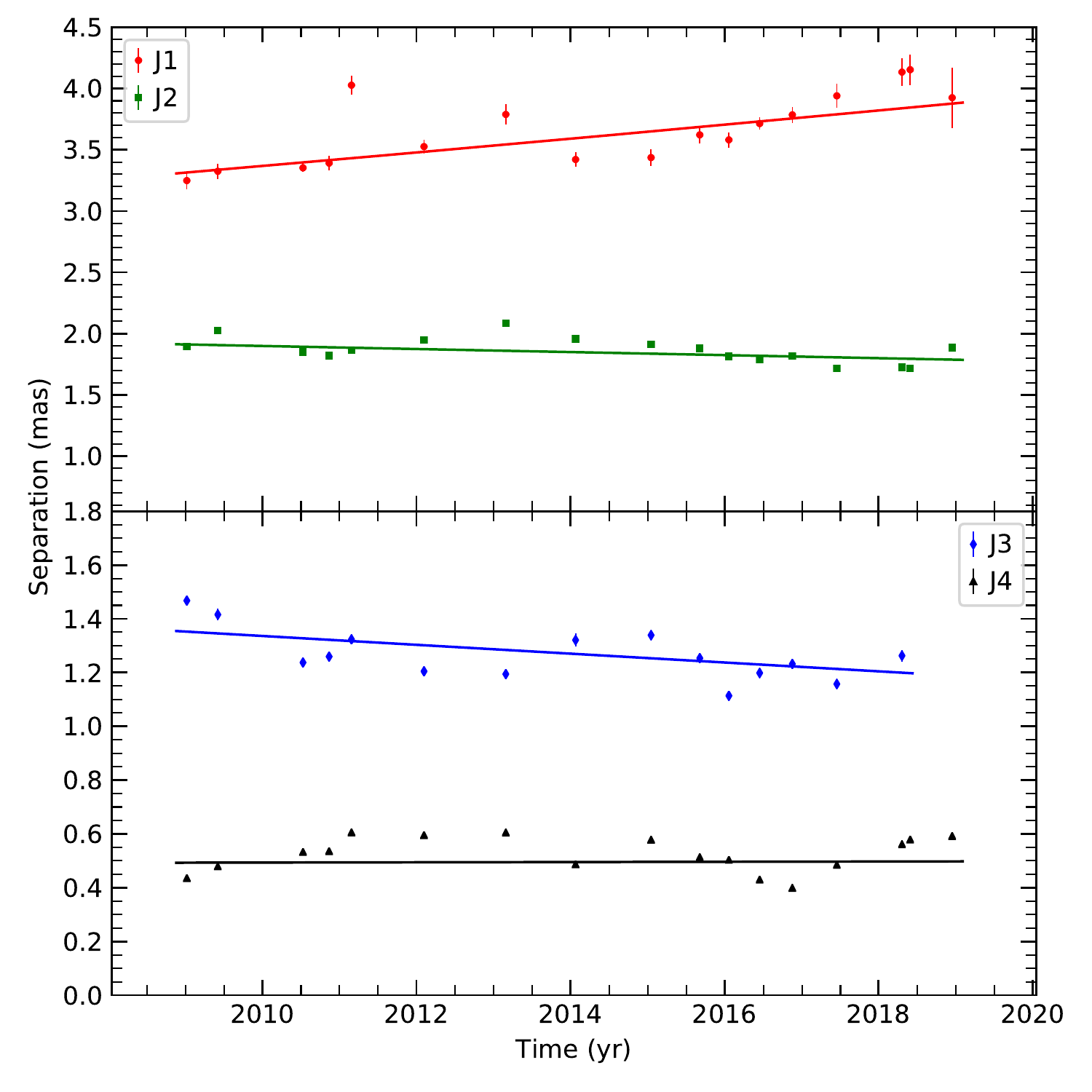}}
\caption{The radial distance of the jet components from the center of the core versus observing time. The data points are derived from 15 GHz multi-epoch observations and presented in Table \ref{tab:comp}. The straight line represents the line regression fitting to the jet proper motion. }
\label{fig3}
\end{figure}

Figure \ref{fig3} shows the radial distance of the jet component from the center of the core versus observing time. A linear regression gives jet component proper motion speeds of 0.071$\pm$0.018 mas yr$^{-1}$ ($1.49\pm0.34\, c$) for J1, and 0.019$\pm$0.003 mas yr$^{-1}$ ($0.41\pm0.06\, c$) for J4 (Table \ref{tab:mu}). It indicates that the apparent jet speed is higher in the outer region (J1: $\sim$90 pc) than the inner portion (J4: $\sim$13 pc). Accounting for the conical jet geometry (Section 3.1), this may not suggest genuine jet acceleration at de-projected distances on tens pc. 
The fitting of J2 and J3 gives negative values, $-0.013\pm0.007$ mas yr$^{-1}$ ($-0.28\pm0.15\, c$) and $-0.017\pm0.007$ mas yr$^{-1}$ ($-0.35\pm0.15\,c$) respectively. 
Since J2 and J3 are located at the jet bending location; therefore their motion is largely affected by the helical trajectory. In the following discussion, we will use the speed estimated for J1 as a representative of the jet speed. Compared with the typical $\gamma$-ray-emitting blazars \citep{2016AJ....152...12L}, the jet proper motion speed of \txs{}, even referring to the fastest jet component J1, is considerably low. 

\begin{deluxetable}{ccccccc}
\tablecaption{Vector Motion Fit Properties of Jet Components \label{tab:mu}}
\tablewidth{0pt}
\tablehead{
\colhead{comp} & \colhead{$N$} & \colhead{$\left<S\right>$} & \colhead{$\left<r\right>$} & \colhead{$\left<\vartheta\right>$} & \colhead{$\mu_{\rm app}$} & \colhead{$\beta_{\rm app}$}\\
\colhead{} & \colhead{} & \colhead{(mJy)} & \colhead{(mas)} & \colhead{($^\circ$)} & \colhead{(mas yr$^{-1}$)} & \colhead{($c$)}
}
\decimalcolnumbers
\startdata
J1 & 17 & 20.8 & 3.669 & 172.1 & 0.071$\pm$0.016 & 1.49$\pm$0.34 \\
J2 & 17 & 46.8 & 1.865 & 188.0 & 0.013$\pm$0.007 & 0.28$\pm$0.15 \\
J3 & 15 & 11.4 & 1.265 & 218.8 & 0.017$\pm$0.007 & 0.35$\pm$0.15 \\
J4 & 17 & 98.2 & 0.524 & 186.6 & 0.019$\pm$0.003 & 0.41$\pm$0.06 
\enddata
\tablecomments{
(1) Component name, (2) number of fitted epochs, (3) mean flux density at 15 GHz, (4) mean distance from core, (5) mean position angle with respect to the core feature,  (6) proper motion, (7) apparent speed in units of the speed of light.
}
\end{deluxetable}

\begin{figure}
\centerline{\includegraphics[width=\linewidth]{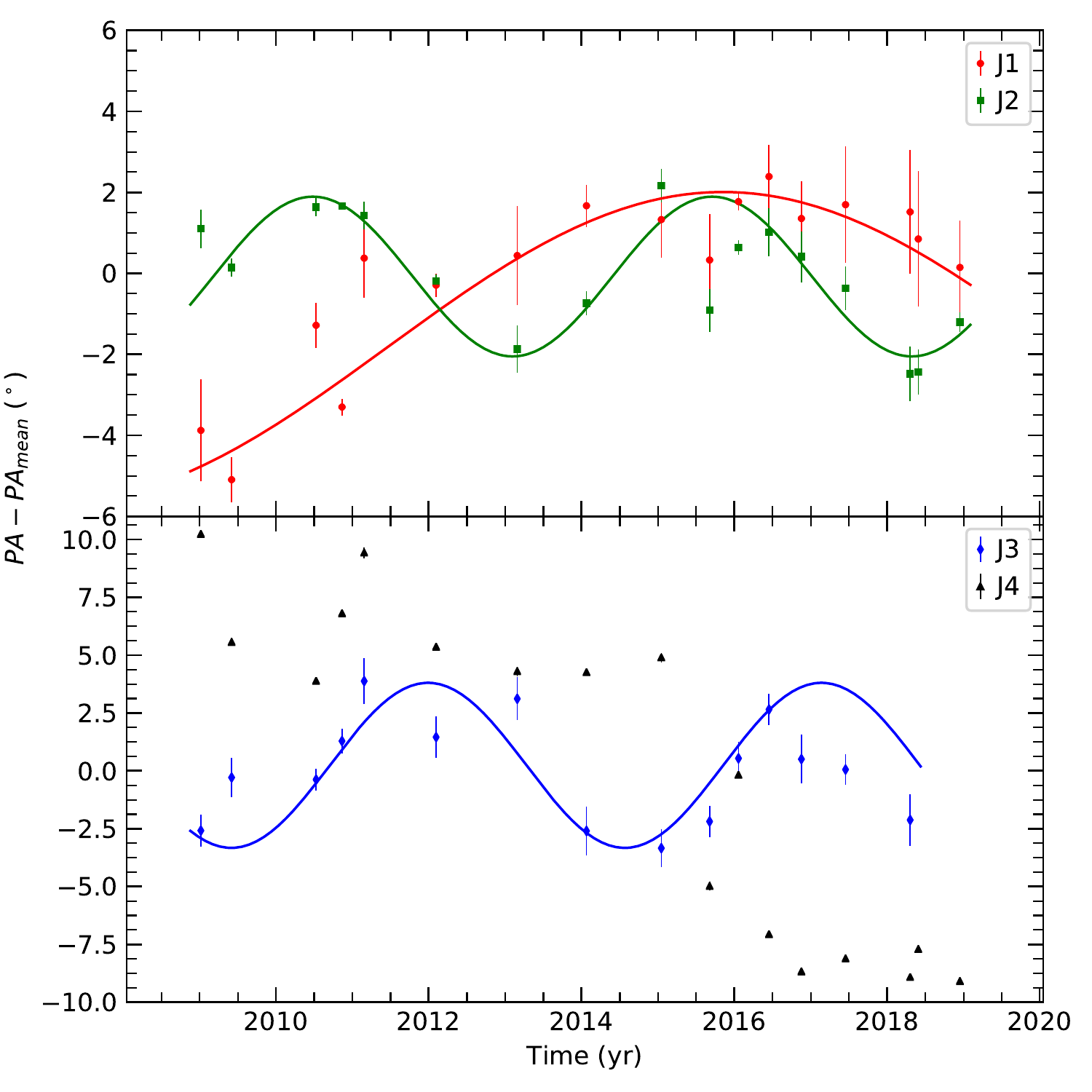}}
\caption{Change of relative position angles of jet components with time.  The data points are derived from 15 GHz multi-epoch observations and presented in Table \ref{tab:comp}. The curves represent the fitting with a sinusoidal function. J2 and J3 exhibit a complete cycle of period of 5.15 yr and 5.23 yr, respectively. J1 also displays significant changes of $dPA$ with time, but does not comprise a complete period, or the scattering is too large to allow for a  good fitting. Similarly, we did not make a fitting to J4 either. }
\label{fig4}
\end{figure}

The east and west boundaries of the jet structure are marked by the components J1 and J3, respectively. A mean position angle (PA) of the jet $PA_{\rm mean} = 195.4\degr$ is estimated by averaging the position angles of J1 and J3. For a scenario involving a precessing motion, such as along a helical trajectory, the jet component position angle swings around $PA_{\rm mean}$. Figure \ref{fig4} shows the relative jet position angles $dPA$ (defined as $dPA = PA - PA_{\rm mean}$) versus the observing time. Components J2 and J3 exhibit this wiggling behaviour, indicated by the oscillations of $dPA$(J2) and $dPA$(J3). A sinusoidal fit to $dPA$ versus time gives periodicities of about 5.15 yr for J2 ($\chi^2 = $ 0.5), and 5.23 yr for J3 ($\chi^2 = $ 0.4). The relative position angle $dPA_{\rm J1}$ ranges between $-5.1\degr$ and $2.4\degr$. When fitting a sinusoidal function with J1, it indicates a period of about 17.61 yr  ($\chi^2 = $ 2.3), but only covers half a cycle. The increasing cycle period with radial distance in the jet components indicates that they trace helical paths. 
The projection of a conical helical line with a constant helix angle parallel to the axis of the cone (here, $PA = 195.4\degr$) onto the plane of the sky is a logarithmic spiral showing increasing pitch with radial distance.
This is consistent with the increasing trend of the apparent jet speed with radial distance.

The component J4 also shows a significant variation, but a periodic pattern is not easily discernible. This is as the model fitting is affected by the brighter core component with J4 in close proximity. 
Whereas, the $dPA$ range of J4, $\pm10\degr$, is substantially larger than those of J1, J2, and J3 which are within $\sim6\degr$, implying that the innermost cone within 0.5 mas (converted to a de-projected physical scale of $\sim$14 pc) may have a larger opening angle. Such a geometry with a wider initial opening angle at the jet base has been noticed before \citep{2020A&A...633L...1R} and has been observed in other AGN \cite[e.g.][]{2018NatAs...2..472G}, and can resemble an inverted funnel structure as expected in kinematic models of the innermost jet \cite[e.g.][]{2015ApJ...805...91M}.

\subsection{Relativistic beaming and jet geometry} \label{sec:pm}

The brightness temperature ($T_{\rm b}$) of a jet component can be employed as an indicator of Doppler boosting. The study of \citet{1969ApJ...155L..71K} first suggested that a maximal brightness temperature of $10^{11} - 10^{12}$ K for a compact self-absorbed radio source may be set by a catastrophic inverse Compton cooling when exceeding $10^{12}$ K. For an incoherent synchrotron radio source, the condition of equipartition of energy between the radiating particles and magnetic fields provides a constraining $T_{\rm b,eq} = 5 \times 10^{10}$~K \citep{1994ApJ...426...51R}.  
Brightness temperature can be calculated from the VLBI observables \citep[see Eq. 4 in ][]{2005AJ....130.2473K}.
The observed brightness temperature $T_{\rm b, obs}$ is amplified compared to the intrinsic $T_{\rm b,int}$ of the synchrotron emission by the Doppler factor $\delta$ when the relativistic jet from the AGN is directed very close to the observer's line of sight. Taking $T_{\rm b,eq}$ as an upper limit of $T_{\rm b,int}$, the Doppler factor is $\delta=T_{\rm b,obs}/T_{\rm b,int}$. The $T_{\rm b,obs}$ and $\delta$ for the \txs{} core component are listed in columns 8--9 in Table \ref{tab:comp}; $\delta$ ranges between 1.6 and 5.8 during 2009 January -- 2016 June (relatively quiescent period) with an average of 2.65.  

The bulk Lorentz factor $\Gamma$ and jet inclination $\theta$ can be calculated from the measurement of jet apparent speed and Doppler boosting factor.
The jet inclination angle is $\theta = 20\pm2\degr$ assuming the above $\beta_{\rm app}$ and $\delta =2.65$ and the intrinsic jet half opening angle is $3.8\pm0.5\degr$. The bulk Lorentz factor during the quiescent period is $\Gamma_{\rm q} = 1.9\pm0.2$. After 2016 November, $\delta$ rises to a maximum of 13.6 on 2017 June 17 followed by a decrease to 9.0 on 2018 December 16. The averaged Doppler factor during the flaring period is $\left<\delta_{\rm f}\right> = 9.8$ with a corresponding $\Gamma_{\rm f} = 5.1\pm0.1$. 
The relatively lower Doppler factor and slower jet speed in \txs{} compared to the $\gamma$-ray emitting blazar population \citep{2016AJ....152...12L} suggests only a moderately relativistic jet in \txs{}.

\subsection{Magnetic field strength and core distance} \label{sec:Bfield}

The frequency-dependent apparent core position shift in blazars is attributable to synchrotron self-absorption in the radio core \citep{1998A&A...330...79L,2005ApJ...619...73H}; it can be deduced through a frequency-dependent time lag in radio flares in single dish measurements \cite[if dominated by the core, e.g.][]{2015MNRAS.452.2004M,2017MNRAS.469..813A} or by an observed shift in the core position from near-simultaneous multi-band VLBI images \cite[e.g.][]{2012A&A...545A.113P,2018ApJ...854...17L} and can be used to estimate the distance of the core from the central engine, the magnetic field strength in the pc-scale jet, jet luminosity and power \cite[e.g.][]{1998A&A...330...79L,2015MNRAS.451..927Z}. Here, we select the multiple frequency VLBI data observed quasi-simultaneously, {\it i.e.}, the separation of observation time is less than two months to minimize effects of variability, to estimate the core shift and magnetic field (Table \ref{tab:coreshift}).

The projected core shift 
\begin{equation}
\Delta r_{\rm proj}=\frac{\Delta d}{2\tan\phi_{\rm proj}}=\frac{\Delta d\sin\theta}{2\tan\phi}
\end{equation}
where $\Delta r_{\rm proj}$ is core shift in mas, $\Delta d=d_1-d_2$ is core size difference between frequency $\nu_1$ and $\nu_2$ in mas, $\phi_{\rm proj}$ and $\phi$ are the projected and real jet half opening angle (in degrees) respectively. The following jet parameters are estimated using the formulation prescribed in \cite{1998A&A...330...79L}. The core shift measure \begin{equation}
\Omega_{r\nu}=4.85\times10^{-9}\frac{D_{L}\Delta r_{\rm proj}}{(1+z)^2(\nu_1^{-1}-\nu_2^{-1})}~{\rm pc~GHz},
\end{equation}
 where $z$ is redshift, $D_L$ is luminosity distance in pc, and $\nu_1$ and $\nu_2$ are observation frequency in GHz. This corresponds to the core distance from the central engine of 
\begin{equation}
r_{\rm core}=\frac{\Omega_{r\nu}}{\nu\sin\theta}.
\end{equation}
The magnetic field strength at a fiducial 1 pc is
\begin{equation}
B_1\cong\left(\frac{\Omega_{r\nu}^3(1+z)^2}{\delta^2\phi\sin^2\theta}\right)^{1/4},
\end{equation}
with a corresponding scaled magnetic field strength in the core $B_{\rm core}=B_{1}(r_{\rm core}/(1~{\rm pc}))^{-1}$. 

We use the redshift $z = 0.3365$, luminosity distance $D_L = 1762$ Mpc, apparent jet speed $\beta_{app} = 1.49$, Doppler factor in the quiescent period $\delta_q = 2.65$ and during the flare $\delta_f = 9.82$, inclination angle $\theta = 20\degr$, and jet half opening angles $\phi_{\rm proj} = 11\degr$ and $\phi = 3.8\degr$. The core shift measure $\Omega_{r\nu}$ ranges between $12.1\pm0.1$ pc GHz to $36.1\pm0.3$ pc GHz (Table \ref{tab:coreshift}). The magnetic field at 1 pc, $B_1$ ranges between ($0.4 \pm 0.1$) G -- ($0.9 \pm 0.2$) G, and that in the core $B_{\rm core}$ ranges between ($0.13 \pm 0.04$) G -- ($0.47 \pm 0.15$) G. The distance of the radio core $r_{\rm core}$ ranges between $0.8 \pm 0.1$ pc to $6.9 \pm 0.7$ pc. 

The magnetic field $B_1$ during the quiescent and flaring periods are $0.69\pm0.03$ G and $0.57\pm0.06$ G respectively, indicating that a portion of the magnetic energy density may have been converted to the radiative energy output during the flare, confined to within the core. The jet parameters and the estimates from the above equations are tabulated in Table \ref{tab:B}.

\subsection{Radio flux density variation} \label{sec:3.5}

Figure \ref{fig5} shows the 15-GHz single-dish radio light curve of \txs{} based on observations made with the OVRO in the top panel. The compiled data points are between 2008 January and 2019 May (about 11.5 yr). The source shows a relatively low-level activity from 2008 to early 2017, with the total flux density ranging between 0.29 Jy to 0.74 Jy and a mean of $0.48\pm 0.1$ Jy. After 2017 September, the source entered a flaring state with the flux density continuing to increase. The flux density is currently larger by a factor of $>3$ compared to the mean value during 2008--2017. This is the largest flare in the observation period of $\sim$11.5 yr. The available data does not allow the identification of the exact peak as it continues to increase. 

The integrated VLBI flux densities (red-colored diamond) at 15 GHz are overlaid onto the OVRO lightcurve (black-colored dots) in the top panel of Figure \ref{fig5}. It shows that the compact VLBI emission dominates the total flux density and follows the same variation trend as the single-dish light curve,  and is consistent with the study of \cite{2019A&A...630A.103B} where this VLBI core dominance was noted. Quantitative comparison between the VLBI flux density and the closest-epoch single-dish one suggests that only 1\%--18\% is attributable to extended emission, which is resolved out in VLBI images. The VLBI core accounts for 53\%--78\% of the integrated VLBI flux density in each individual epoch (see Table \ref{tab:comp} and the middle panel of Figure \ref{fig5}) and also shows a significantly increasing trend since 2017. A moderate level of variability is found in jet components J1, J2, and J3. The innermost jet component J4 shows notable variation, with a minimum of 45 mJy on 2013 February 28 and a maximum of 154 mJy on 2016 November 18. This may largely owe to rapid local changes in the inclination angle and the corresponding Doppler beaming factor, such as in a turbulent environment or along a helical trajectory.

As the radio flux density is mainly dominated by the core, the aperiodic variability may be attributed to episodes of injection of energetic particles at the jet base. This can result in increased flux density from relativistically beamed propagating shock fronts \citep{1979ApJ...232...34B}. Equating the variability timescale $t_{\rm var}$ to the synchrotron cooling timescale, 
\begin{equation}
\frac{\delta t_{\rm var}}{1+z} = \frac{3 m_e c}{4 \sigma_T \beta^2 \gamma U_B} 
\end{equation}
where $m_e = 9.11 \times 10^{-28}$ g is the electron mass, $\sigma_T = 6.65 \times 10^{-25}$ cm$^2$ is the Thomson cross section, $\beta \approx 1$ is the velocity and $\gamma \approx 100$ is the Lorentz factor of the injected electrons, and $U_B = B^2/(8 \pi)$ is the magnetic field energy density. Using $t_{\rm var} = 240$ days  obtained as the average flare duration from Table 1 of \cite{2019MNRAS.483L..42K}, the size of the emission region $\Delta r \leq \delta t_{var}/(1+z) = 0.3$ pc assuming that $\delta = \delta_q = 2.65$ is representative of the true size (during the quiescent period). This corresponds to a distance $r = \Delta r/\sin \phi = 4.5$ pc which is an upper limit but similar to the typical $r_{\rm core}$. The magnetic field strength is $B = 0.7~{\rm G}~\delta^{-1/2}$ which is 0.43 G during the quiescent period and 0.23 G during the flaring period marking a more stark contrast. These values are nearly similar to the core-shift estimates, and suggest a conversion of magnetic energy density to particle kinetic energy which then helps accelerate the particles upon injection at the jet base.

The helical geometry at larger scales is suggestive of an origin from instabilities in the jet. This has been suggested as a likely mechanism in the blazars 1156$+$295 \citep{2011A&A...529A.113Z}, 3C 48 \citep{2010MNRAS.402...87A} and 3C 273 \citep{2006A&A...456..493P} where the Kelvin-Helmholtz instability operational from the sub-pc scales causes the helically structured jet. This scenario may likely be occurring in \txs{} as well, since the jet launching region does not exhibit periodic signatures, though this requires further high-resolution VLBI monitoring observations to confirm.

The neutrino and associated $\gamma$-ray flares were discovered in the early rising phase of the flare \citep{2018Sci...361.1378I}, marked with a vertical blue line in Figure \ref{fig5}. Another period, in which a 3.5$\sigma$ neutrino event was detected, is indicated with a red-colored shade zone. It is puzzling that the 2017 neutrino flare occurred in the still-rising phase of the radio flux density; after the neutrino event, the flux density keeps rising to a maximum of 1.55 Jy on 2019 April 25. It is possible that the neutrino event and $\gamma$-ray flares were associated with the onset of the injection of energetic particles into the optically thick jet base.

\begin{deluxetable}{ccccccccccc}
\tablecaption{Core shift measurements. \label{tab:coreshift}}
\tablewidth{0pt}
\tablehead{
\colhead{Date 1} & \colhead{$\nu_1$} & \colhead{$d_1$} & \colhead{Date 2} & \colhead{$\nu_2$} & \colhead{$d_2$} & \colhead{$\Delta{}r$} & \colhead{$\Omega_{r\nu}$} & \colhead{$B_\mathrm{1pc}$} & \colhead{$r_\mathrm{core}$} & \colhead{$B_\mathrm{core}$} \\
\colhead{(Y-M-D)} & \colhead{(GHz)} & \colhead{(mas)} & \colhead{(Y-M-D)} & \colhead{(GHz)} & \colhead{(mas)} & \colhead{(mas)} & \colhead{(pc\,GHz)} & \colhead{(G)} & \colhead{(pc)} & \colhead{(G)}
}
\decimalcolnumbers
\startdata
2003-09-13 & 24.4 & 0.107$\pm$0.003 & 2003-09-13 & 43.1 & 0.089$\pm$0.003 & 0.017$\pm$0.006 & 12.1$\pm$0.1 & 0.4$\pm$0.1 & 0.8$\pm$0.1 & 0.47$\pm$0.15 \\
2010-10-23 & 8.6 & 0.229$\pm$0.001 & 2010-11-13 & 15.4 & 0.081$\pm$0.001 & 0.148$\pm$0.002 & 36.1$\pm$0.3 & 0.9$\pm$0.2 & 6.9$\pm$0.7 & 0.13$\pm$0.04 \\
2011-02-07 & 8.4 & 0.234$\pm$0.001 & 2011-02-27 & 15.4 & 0.105$\pm$0.001 & 0.129$\pm$0.002 & 29.2$\pm$0.3 & 0.8$\pm$0.2 & 5.6$\pm$0.6 & 0.14$\pm$0.04 \\
2012-03-15 & 8.4 & 0.220$\pm$0.003 & 2012-02-06 & 15.4 & 0.077$\pm$0.001 & 0.143$\pm$0.004 & 32.3$\pm$0.3 & 0.8$\pm$0.2 & 6.2$\pm$0.7 & 0.13$\pm$0.04 \\
2013-02-12 & 7.6 & 0.218$\pm$0.003 & 2013-02-28 & 15.4 & 0.109$\pm$0.001 & 0.109$\pm$0.004 & 20.3$\pm$0.4 & 0.6$\pm$0.1 & 3.9$\pm$0.5 & 0.15$\pm$0.05 \\
2015-10-07 & 7.6 & 0.256$\pm$0.002 & 2015-09-06 & 15.4 & 0.120$\pm$0.001 & 0.136$\pm$0.003 & 25.4$\pm$0.4 & 0.7$\pm$0.1 & 4.8$\pm$0.5 & 0.14$\pm$0.05 \\
2016-06-28 & 7.6 & 0.203$\pm$0.001 & 2016-06-16 & 15.4 & 0.087$\pm$0.001 & 0.116$\pm$0.002 & 21.6$\pm$0.4 & 0.6$\pm$0.1 & 4.1$\pm$0.5 & 0.15$\pm$0.05 \\
2018-07-21 & 8.7 & 0.179$\pm$0.001 & 2018-05-31 & 15.4 & 0.097$\pm$0.001 & 0.083$\pm$0.001 & 20.2$\pm$0.3 & 0.6$\pm$0.1 & 3.8$\pm$0.4 & 0.15$\pm$0.05
\enddata
\tablecomments{{ 
(1)--(3) Observation date, frequency, and core size of the low frequency data; (4)--(6) Observation date, frequency, and core size of the high frequency data; (7) Core shift; (8) Core shift measure; (9) magnetic field at 1 pc of actual distance from the jet vertex; (10) Absolute distance of VLBI core from jet vertex; (11) Magnetic field strength of VLBI core.}
}
\end{deluxetable}

\begin{table}
    \caption{Magnetic field strength before and during 2017 flare} 
    \centering
    \begin{tabular}{c|c} \hline\hline
Parameter & Values \\ \hline
$z$ & 0.3365 \\
$D_L$ & 1762 Mpc \\
$\beta_{app}$ & $1.49\pm0.34$ \\
$\left<\delta\right>$ & $2.65\pm0.01$ \\
$\left<\Gamma_q\right>$ & $1.9\pm0.2$\\
$\left<\theta\right>$ & $20\pm2\degr$\\
$\phi_{proj}$ & $11.0\pm0.3\degr$ \\
$\phi_{int}$ & $3.8\pm0.5\degr$\\
$B_{1pc}^{before}$ & $0.69\pm0.03$ G\\
$B_{1pc}^{during}$ & $0.57\pm0.06$ G\\   \hline
\end{tabular}
    \label{tab:B}
\tablecomments{$z$: red shift; $D_L$: luminosity distance; $\beta_{app}$: the apparent speed in units of the speed of light; $\left<\delta\right>$ and $\left<\Gamma_q\right>$: the mean Doppler factor and bulk Lorentz factor during the quiescent period, respectively; $\left<\theta\right>$: the jet inclination angle; $\phi_{proj}$: the apparent half opening angle; $\phi_{int}$: the intrinsic jet half opening angle; $B_{1pc}^{before}$ and $B_{1pc}^{during}$: magnetic field at 1 pc of actual distance from the jet vertex of quiescent and flaring periods respectively.}
\end{table}

\section{Summary}\label{summary} \label{sec:5}

The pc-scale jet properties of the neutrino--emitting blazar \txs{} are explored here using multi-frequency, multi-epoch VLBI data. The jet morphology and evolution suggest a helical jet structure originating in growing instabilities. The oscillation of the jet beam position angle indicates a periodicity of 5--6 yr. The jet geometry is characterized by an inclination angle of $\sim20\degr$ and a half opening angle of $\sim3.8\degr$. The apparent jet speed, Doppler boosting factor and bulk Lorentz factor point towards a moderately relativistic jet. The magnetic field strength at pc-scales is found to decrease during the recent ongoing flaring period compared to that during the quiescent period. This suggests a conversion of the magnetic field energy density to particle energy density, which helps accelerate particles upon injection at the jet base which subsequently produces the flares. 

The aperiodic radio light curve, dominated by the core, suggests that the periodic PA variations of the components are confined to the outer region, thus implying an origin in growing jet instabilities. Further, the 15 GHz as well as the higher frequency images do not indicate a secondary radio bright core component at the pc-scale in addition to the identified core. This provides less evidence for a scenario involving a SMBH binary causing a precessing jet either through tidal torques or a twin jet through Lense-Thirring precession of the inner accretion disk \citep{2019A&A...630A.103B}. However, we cannot rule out an SMBH binary entirely owing to the possibility that the secondary SMBH may not be an active radio emitter or may be radio quiet \cite[e.g.][]{2018RaSc...53.1211A}. Scenarios involving a successful SMBH merger driving a relativistic jet \citep{2019MNRAS.483L..42K} or a spine-sheath structured jet \citep{2020A&A...633L...1R} produce high-energy cosmic rays and particle decays; the emergent picture in our study is consistent with this owing to it involving injection events at the jet base.

The neutrino event detected during the rise of the radio flare could then be associated with the onset of particle injection and acceleration which can contain a large portion of the converted energy density. This scenario provides support to a lepto-hadronic origin of the VHE neutrinos and $\gamma$-ray emission owing to a co-spatial origin at the particle injection and acceleration site.
The available archive VLBI data are however far separated in time from the neutrino--$\gamma$-flare, not allowing for stronger constraints on the physical conditions during the 2017 flare. 
With the radio flux density of \txs{} still continuing to rise, ongoing multi-wavelength monitoring would be necessary to reveal the exact peak time of this prominent flare. 
An accurate identification of the peak time can help improve the modeling of the core shift estimates by using the time delays between peak times at different frequencies, as compared to our current estimate based on the VLBI imaging.
This can additionally improve the estimated variability timescale (which currently does not include this epoch), helpful in the calculation and confirmation of the pc-scale magnetic field strength. 
These observations may also provide useful information should there be future detections of neutrino events from \txs{} again.

\section{Acknowledgments}\label{sec:6}

The authors are grateful to the anonymous referee for his/her helpful comments. This work is supported by the National Key R\&D Programme of China (grant number 2018YFA0404602, 2018YFA0404603), Chinese Academy of Sciences ( grant number 114231KYSB20170003), and National Natural Science Foundation of China (Grant No. 11803062). XFL thanks Shahzad Anjum for the discussion about the SED.
This research has made use of data from the MOJAVE database that is maintained by the MOJAVE team \citep{2009AJ....137.3718L}. This research has made use of the NASA/IPAC Extragalactic Database (NED), which is operated by the Jet Propulsion Laboratory, California Institute of Technology, under contract with the National Aeronautics and Space Administration. This research has made use of data from the OVRO 40-m monitoring program \citep{2011ApJS..194...29R} which is supported in part by NASA grants NNX08AW31G, NNX11A043G, and NNX14AQ89G and NSF grants AST-0808050 and AST-1109911.

\vspace{5mm}
\facilities{VLBA(NRAO), OVRO 40m}

\software{AIPS   \citep{2003ASSL..285..109G},  
          DIFMAP \citep{1997ASPC..125...77S}
          }

\clearpage

\appendix 

\section{VLBI Data Analysis}
 \label{sec:A1}

The 8 GHz VLBI data was obtained from the VLBI Calibrator Survey (VCS) project \footnote{Astrogeo VLBI FITS image database maintained by Leonid Petrov: \url{http://astrogeo.org/}} and was calibrated with the PIMA software package \citep{2011AJ....142...35P}. The 15 GHz data was acquired from the Monitoring Of Jets in Active galactic nuclei with the Very long baseline array (VLBA) Experiments \citep[MOJAVE; ][]{2009AJ....137.3718L} survey archive and has been processed with the Astronomical Image Processing System (AIPS) software package \citep{2003ASSL..285..109G}.
The 24 and 43 GHz data are from high-frequency celestial reference frame \citep[CRF;][]{2010AJ....139.1695L} which were obtained from the National Radio Astronomy Observatory (NRAO) data archive\footnote{\href{https://archive.nrao.edu/archive/advquery.jsp}{https://archive.nrao.edu/archive/advquery.jsp}} and calibrated using Astronomical Image Processing System (AIPS) by running appropriate tasks and procedures. The standard VLBI calibration was applied using the AIPS software.

All the calibrated VLBI data were imported to {\sc DIFMAP} in order to carry out phase and amplitude self-calibration. After some iterations of self-calibration, the noise in the residual image is rendered close to the thermal noise. Then we fit the data with several Gaussian models in {\sc DIFMAP} to quantitatively describe the emission properties of the VLBI components. The model fitting parameters, including the integrated flux density ($S_i$), radial separation ($r$) with respect to the core and the position angle (PA), and full width half maximum component size $d$, are listed in Table~\ref{tab:comp}. When the component is too compact to be resolved, the Gaussian model size is reduced to a point source. In this case, an upper limit of the component size is estimated by using the method presented in \citet{2005astro.ph..3225L}. 
\begin{equation}
d_{\rm min} = \frac{2^{1+\beta/2}}{\pi} \left[\pi B_{\rm maj}B_{\rm min}\ln 2 \ln\frac{SNR}{SNR-1}\right]^{1/2} \\
\end{equation}
where $SNR=\frac{S_{\rm peak}}{\sigma_{\rm rms}}$, $\beta=2$ for uniform weighting or $\beta=0$ for natural weighting. A natural weighting scheme was used to produce all the images in this paper.
We adopted 5\% amplitude calibration error as the flux density uncertainty. The uncertainties of the component distance $r$ and size $d$ are estimated by taking into account the signal-to-noise ratio \citep[][]{1999ASPC..180..301F}.
\begin{align}
\Delta S_{\rm tot} &= 5\% S_{\rm tot} \nonumber \\
\Delta d &= d / SNR \nonumber \\
\Delta x &= \frac{d}{2} / SNR \nonumber \\
\Delta y &= \frac{d}{2} / SNR \nonumber \\
r &= \sqrt{x^2+y^2} \\
PA &= \arctan\left(\frac{x}{y}\right) \nonumber \\
\Delta r &= \frac{\left|x\right|}{r}\Delta x + \frac{\left| y\right|}{r} \Delta y \nonumber \\
\Delta P{}A &= \frac{\left|y\right|}{x^2+y^2}\Delta x + \frac{\left|x\right|}{x^2+y^2}\Delta y \nonumber
\end{align}
The observational parameters are summarized in Table~\ref{tab:obs}.
\startlongtable
\begin{deluxetable}{ccccccccc}
\tablecaption{Observation Logs. \label{tab:obs}}
\tablewidth{0pt}
\tablehead{
\colhead{Date} & \colhead{Code} & \colhead{$\nu$} & \colhead{$BW$} & \colhead{$t_{int}$} & \colhead{$S_{int}$} & \colhead{$S_{peak}$} & \colhead{$rms$} & \colhead{$Beam$} \\ 
\colhead{(Y-M-D)} & \colhead{} & \colhead{(GHz)} & \colhead{(MHz)} & \colhead{(min)} & \colhead{(mJy)} & \colhead{(mJy/beam)} & \colhead{(mJy/beam)} & \colhead{($\mathrm{mas}\times\mathrm{mas},^\circ$)}
}
\decimalcolnumbers
\startdata
2003-09-13 & BL115 & 24.4 & 64 & 5 & 406.9 & 297.7 & 1.9 & 0.90$\times$0.33, $-$13.3 \\
2003-09-13 & BL115 & 43.1 & 64 & 8 & 250.1 & 182.0 & 1.9 & 0.58$\times$0.20, $-$12.1 \\
2009-01-07 & BL149 & 15.4 & 64 & 45 & 544.1 & 417.8 & 0.2 & 1.26$\times$0.59, $-$2.4 \\
2009-06-03 & BL149 & 15.4 & 64 & 36 & 591.9 & 440.9 & 0.2 & 1.26$\times$0.58, $-$3.7 \\
2010-07-12 & BL149CL & 15.4 & 64 & 36 & 423.2 & 276.2 & 0.2 & 1.32$\times$0.51, $-$6.7 \\
2010-10-23 & BC191N & 8.6 & 128 & 5 & 374.6 & 246.5 & 0.5 & 2.05$\times$0.81, $-$2.9 \\
2010-11-13 & BL149CW & 15.4 & 64 & 36 & 378.0 & 245.7 & 0.2 & 1.23$\times$0.56, $-$5.9 \\
2011-02-07 & BC196F & 8.4 & 128 & 13 & 394.2 & 272.4 & 0.3 & 2.20$\times$0.96,    4.5 \\
2011-02-27 & BL149DC & 15.4 & 64 & 34 & 357.8 & 232.6 & 0.3 & 1.13$\times$0.53,    1.2 \\
2012-02-06 & BL178AG & 15.4 & 64 & 38 & 346.8 & 235.7 & 0.2 & 1.20$\times$0.55, $-$3.3 \\
2012-03-15 & BC201AR & 8.4 & 128 & 1 & 362.3 & 265.8 & 0.6 & 2.17$\times$0.94,    6.2 \\
2013-02-12 & BP171A2 & 7.6 & 256 & 1 & 280.4 & 222.0 & 0.7 & 2.14$\times$0.84, $-$2.6 \\
2013-02-28 & BL178BA & 15.4 & 64 & 34 & 341.1 & 237.4 & 0.2 & 1.17$\times$0.54, $-$2.9 \\
2014-01-25 & BL193AE & 15.4 & 256 & 50 & 423.5 & 311.8 & 0.2 & 1.16$\times$0.56, $-$3.7 \\
2015-01-18 & BL193AQ & 15.4 & 256 & 53 & 422.0 & 300.5 & 0.2 & 1.25$\times$0.52, $-$2.7 \\
2015-09-06 & BL193AW & 15.4 & 256 & 51 & 386.5 & 271.4 & 0.1 & 1.18$\times$0.50, $-$4.1 \\
2015-10-07 & BP192B3 & 7.6 & 256 & 3 & 314.3 & 239.6 & 0.2 & 2.38$\times$1.09,    5.7 \\
2016-01-22 & BL193BB & 15.4 & 256 & 55 & 311.5 & 210.9 & 0.2 & 1.21$\times$0.53, $-$6.5 \\
2016-06-16 & BL193BG & 15.4 & 256 & 51 & 435.8 & 319.0 & 0.2 & 1.14$\times$0.51, $-$0.6 \\
2016-06-28 & S7104A1 & 7.6 & 256 & 1 & 343.5 & 268.1 & 0.3 & 2.37$\times$1.05, $-$1.6 \\
2016-11-18 & BL193BM & 15.4 & 256 & 49 & 530.2 & 404.3 & 0.2 & 1.15$\times$0.53, $-$2.9 \\
2017-06-17 & BL229AI & 15.4 & 256 & 35 & 591.6 & 453.4 & 0.3 & 1.15$\times$0.50, $-$4.7 \\
2018-04-22 & BL229AN & 15.4 & 256 & 37 & 894.3 & 712.1 & 0.3 & 1.17$\times$0.52, $-$4.1 \\
2018-05-31 & BL229AO & 15.4 & 256 & 35 & 977.9 & 794.7 & 0.4 & 1.15$\times$0.50, $-$2.4 \\
2018-07-21 & BP222A & 8.7 & 384 & 3 & 898.5 & 755.2 & 0.4 & 2.16$\times$0.87, $-$4.8 \\
2018-12-16 & BL229AT & 15.4 & 256 & 29 &1018.2 & 842.1 & 0.5 & 1.19$\times$0.56,    3.8 \\
\enddata
\tablecomments{
(1) Observation date; 
(2) Project code; 
(3) Frequency; 
(4) Bandwidth; 
(5) Integration time; 
(6) Integrated flux density; 
(7) Peak flux density; 
(8) {\it rms} noise of the image; 
(9) Major axis, minor axis, and position angle (measured from North to East) of the major axis of the restoring beam major axis.
	}
\end{deluxetable}

\startlongtable
\begin{deluxetable}{ccccccccc}
\tablecaption{Fitted parameters of VLBI components. \label{tab:comp}}
\tablewidth{0pt}
\tablehead{
\colhead{Date} & \colhead{$\nu$} & \colhead{Comp} & \colhead{$S$} & \colhead{$r$} & \colhead{$PA$} & \colhead{$d$} & \colhead{$T_{\rm b,obs}$} & \colhead{$\delta$}\\
\colhead{(Y-M-D)} & \colhead{(GHz)} & \colhead{} & \colhead{(mJy)} & \colhead{(mas)} & \colhead{($^\circ$)} & \colhead{(mas)} & \colhead{($10^{11}$ K)} & \colhead{}
}
\decimalcolnumbers
\startdata
2003-09-13 & 24.4 & C & 306.6$\pm$4.5 & 0 & 0 & 0.107$\pm$0.003 & 0.73 & 1.46 \\
2003-09-13 & 43.1 & C & 204.2$\pm$5.0 & 0 & 0 & 0.089$\pm$0.003 & 0.22 & 0.45 \\
2009-01-07 & 15.4 & C & 354.9$\pm$0.9 & 0 & 0 & 0.092$\pm$0.001 & 2.91 & 5.81 \\
 & & J4 & 101.8$\pm$0.70 & 0.435$\pm$0.001 & $-$163.2$\pm$0.1 & 0.199$\pm$0.001 & & \\
 & & J3 & 7.8$\pm$0.60 & 1.467$\pm$0.018 & $-$143.8$\pm$0.5 & 0.108$\pm$0.015 & & \\
 & & J2 & 46.2$\pm$1.35 & 1.893$\pm$0.016 & $-$170.9$\pm$0.1 & 0.804$\pm$0.011 & & \\
 & & J1 & 24.7$\pm$2.90 & 3.249$\pm$0.072 & 168.2$\pm$0.3 & 1.753$\pm$0.050 & & \\
2009-06-03 & 15.4 & C & 369.7$\pm$0.8 & 0 & 0 & 0.114$\pm$0.001 & 1.98 & 3.97 \\
 & & J4 & 131.3$\pm$0.53 & 0.479$\pm$0.001 & $-$167.8$\pm$0.0 & 0.290$\pm$0.001 & & \\
 & & J3 & 11.3$\pm$0.70 & 1.416$\pm$0.021 & $-$141.5$\pm$0.7 & 0.352$\pm$0.017 & & \\
 & & J2 & 48.7$\pm$1.60 & 2.025$\pm$0.019 & $-$171.8$\pm$0.1 & 0.897$\pm$0.013 & & \\
 & & J1 & 23.7$\pm$2.42 & 3.325$\pm$0.063 & 167.0$\pm$0.3 & 1.877$\pm$0.043 & & \\
2010-07-12 & 15.4 & C & 221.5$\pm$0.5 & 0 & 0 & 0.096$\pm$0.001 & 1.68 & 3.35 \\
 & & J4 & 108.6$\pm$0.34 & 0.532$\pm$0.001 & $-$169.5$\pm$0.0 & 0.302$\pm$0.001 & & \\
 & & J3 & 23.1$\pm$0.50 & 1.237$\pm$0.009 & $-$141.6$\pm$0.3 & 0.497$\pm$0.006 & & \\
 & & J2 & 42.8$\pm$0.65 & 1.850$\pm$0.008 & $-$170.3$\pm$0.1 & 0.738$\pm$0.005 & & \\
 & & J1 & 23.3$\pm$1.28 & 3.353$\pm$0.035 & 170.8$\pm$0.2 & 1.594$\pm$0.022 & & \\
2010-10-23 & 8.6 & C & 223.4$\pm$0.6 & 0 & 0 & 0.229$\pm$0.001 & 0.93 & 1.86 \\
2010-11-13 & 15.4 & C & 212.4$\pm$0.9 & 0 & 0 & 0.081$\pm$0.001 & 2.27 & 4.53 \\
 & & J4 & 67.3$\pm$0.57 & 0.535$\pm$0.002 & $-$166.6$\pm$0.1 & 0.218$\pm$0.001 & & \\
 & & J3 & 17.8$\pm$0.87 & 1.259$\pm$0.015 & $-$139.9$\pm$0.6 & 0.328$\pm$0.012 & & \\
 & & J2 & 56.0$\pm$1.47 & 1.821$\pm$0.014 & $-$170.3$\pm$0.1 & 0.816$\pm$0.009 & & \\
 & & J1 & 19.3$\pm$1.91 & 3.391$\pm$0.060 & 168.8$\pm$0.3 & 1.695$\pm$0.040 & & \\
2011-02-07 & 8.4 & C & 249.3$\pm$0.5 & 0 & 0 & 0.234$\pm$0.001 & 1.06 & 2.12 \\
2011-02-27 & 15.4 & C & 217.6$\pm$0.8 & 0 & 0 & 0.105$\pm$0.001 & 1.36 & 2.72 \\
 & & J4 & 55.6$\pm$0.72 & 0.605$\pm$0.003 & $-$164.0$\pm$0.1 & 0.303$\pm$0.002 & & \\
 & & J3 & 8.4$\pm$0.66 & 1.323$\pm$0.017 & $-$137.3$\pm$0.7 & 0.083$\pm$0.016 & & \\
 & & J2 & 58.8$\pm$1.44 & 1.863$\pm$0.012 & $-$170.6$\pm$0.1 & 0.913$\pm$0.009 & & \\
 & & J1 & 15.4$\pm$2.16 & 4.028$\pm$0.077 & 172.5$\pm$0.2 & 1.408$\pm$0.053 & & \\
2012-02-06 & 15.4 & C & 217.0$\pm$0.6 & 0 & 0 & 0.077$\pm$0.001 & 2.55 & 5.09 \\
 & & J4 & 45.1$\pm$0.35 & 0.595$\pm$0.002 & $-$168.1$\pm$0.0 & 0.210$\pm$0.001 & & \\
 & & J3 & 10.7$\pm$0.68 & 1.205$\pm$0.020 & $-$139.8$\pm$0.8 & 0.386$\pm$0.017 & & \\
 & & J2 & 54.4$\pm$1.83 & 1.948$\pm$0.019 & $-$172.2$\pm$0.1 & 0.960$\pm$0.013 & & \\
 & & J1 & 17.0$\pm$1.69 & 3.526$\pm$0.058 & 171.8$\pm$0.2 & 1.602$\pm$0.040 & & \\
2012-03-15 & 8.4 & C & 259.7$\pm$1.6 & 0 & 0 & 0.220$\pm$0.003 & 1.26 & 2.51 \\
2013-02-12 & 7.6 & C & 211.1$\pm$1.3 & 0 & 0 & 0.218$\pm$0.003 & 1.25 & 2.49 \\
2013-02-28 & 15.4 & C & 222.9$\pm$0.8 & 0 & 0 & 0.109$\pm$0.001 & 1.29 & 2.58 \\
 & & J4 & 44.9$\pm$0.54 & 0.604$\pm$0.002 & $-$169.1$\pm$0.0 & 0.298$\pm$0.001 & & \\
 & & J3 & 8.5$\pm$0.51 & 1.194$\pm$0.019 & $-$138.1$\pm$0.8 & 0.445$\pm$0.017 & & \\
 & & J2 & 44.3$\pm$1.71 & 2.084$\pm$0.021 & $-$173.8$\pm$0.1 & 1.056$\pm$0.015 & & \\
 & & J1 & 17.2$\pm$2.41 & 3.789$\pm$0.081 & 172.5$\pm$0.2 & 1.819$\pm$0.055 & & \\
2014-01-25 & 15.4 & C & 283.9$\pm$0.9 & 0 & 0 & 0.139$\pm$0.001 & 1.01 & 2.02 \\
 & & J4 & 69.0$\pm$0.69 & 0.486$\pm$0.002 & $-$169.2$\pm$0.1 & 0.295$\pm$0.001 & & \\
 & & J3 & 9.4$\pm$0.67 & 1.321$\pm$0.024 & $-$143.8$\pm$0.8 & 0.666$\pm$0.020 & & \\
 & & J2 & 33.6$\pm$1.24 & 1.957$\pm$0.019 & $-$172.7$\pm$0.1 & 1.030$\pm$0.013 & & \\
 & & J1 & 22.6$\pm$2.41 & 3.421$\pm$0.061 & 173.8$\pm$0.2 & 2.061$\pm$0.042 & & \\
2015-01-18 & 15.4 & C & 283.0$\pm$1.1 & 0 & 0 & 0.156$\pm$0.001 & 0.80 & 1.61 \\
 & & J4 & 66.4$\pm$0.77 & 0.578$\pm$0.002 & $-$168.5$\pm$0.1 & 0.318$\pm$0.002 & & \\
 & & J3 & 10.1$\pm$0.50 & 1.339$\pm$0.016 & $-$144.6$\pm$0.5 & 0.560$\pm$0.013 & & \\
 & & J2 & 33.6$\pm$1.00 & 1.912$\pm$0.016 & $-$169.8$\pm$0.1 & 0.863$\pm$0.010 & & \\
 & & J1 & 23.2$\pm$2.54 & 3.437$\pm$0.067 & 173.4$\pm$0.2 & 1.926$\pm$0.043 & & \\
2015-09-06 & 15.4 & C & 238.7$\pm$0.9 & 0 & 0 & 0.120$\pm$0.001 & 1.15 & 2.31 \\
 & & J4 & 75.8$\pm$0.63 & 0.513$\pm$0.002 & $-$178.4$\pm$0.0 & 0.340$\pm$0.001 & & \\
 & & J3 & 11.1$\pm$0.53 & 1.253$\pm$0.015 & $-$143.4$\pm$0.5 & 0.377$\pm$0.011 & & \\
 & & J2 & 37.1$\pm$1.20 & 1.880$\pm$0.017 & $-$172.9$\pm$0.1 & 0.955$\pm$0.011 & & \\
 & & J1 & 19.3$\pm$2.40 & 3.622$\pm$0.072 & 172.4$\pm$0.2 & 1.682$\pm$0.047 & & \\
2015-10-07 & 7.6 & C & 223.3$\pm$0.9 & 0 & 0 & 0.256$\pm$0.002 & 0.96 & 1.92 \\
2016-01-22 & 15.4 & C & 172.5$\pm$0.5 & 0 & 0 & 0.076$\pm$0.001 & 2.07 & 4.14 \\
 & & J4 & 68.6$\pm$0.44 & 0.503$\pm$0.002 & $-$173.6$\pm$0.0 & 0.268$\pm$0.001 & & \\
 & & J3 & 11.4$\pm$0.54 & 1.114$\pm$0.016 & $-$140.7$\pm$0.7 & 0.502$\pm$0.013 & & \\
 & & J2 & 36.3$\pm$1.46 & 1.815$\pm$0.022 & $-$171.3$\pm$0.2 & 0.921$\pm$0.015 & & \\
 & & J1 & 20.6$\pm$2.13 & 3.581$\pm$0.062 & 173.9$\pm$0.2 & 1.730$\pm$0.041 & & \\
2016-06-16 & 15.4 & C & 254.9$\pm$0.5 & 0 & 0 & 0.087$\pm$0.000 & 2.32 & 4.64 \\
 & & J4 & 103.7$\pm$0.47 & 0.430$\pm$0.001 & 179.5$\pm$0.0 & 0.294$\pm$0.001 & & \\
 & & J3 & 10.6$\pm$0.49 & 1.198$\pm$0.014 & $-$138.6$\pm$0.6 & 0.419$\pm$0.012 & & \\
 & & J2 & 42.7$\pm$1.52 & 1.788$\pm$0.019 & $-$171.0$\pm$0.1 & 0.942$\pm$0.013 & & \\
 & & J1 & 24.2$\pm$2.20 & 3.714$\pm$0.051 & 174.5$\pm$0.1 & 1.802$\pm$0.034 & & \\
2016-06-28 & 7.6 & C & 252.4$\pm$0.6 & 0 & 0 & 0.203$\pm$0.001 & 1.72 & 3.44 \\
2016-11-18 & 15.4 & C & 304.0$\pm$0.5 & 0 & 0 & 0.077$\pm$0.000 & 3.52 & 7.05 \\
 & & J4 & 154.2$\pm$0.43 & 0.398$\pm$0.001 & 177.9$\pm$0.0 & 0.315$\pm$0.000 & & \\
 & & J3 & 8.9$\pm$0.58 & 1.232$\pm$0.019 & $-$140.7$\pm$0.7 & 0.305$\pm$0.016 & & \\
 & & J2 & 38.5$\pm$1.53 & 1.817$\pm$0.021 & $-$171.6$\pm$0.1 & 0.919$\pm$0.014 & & \\
 & & J1 & 22.7$\pm$2.59 & 3.785$\pm$0.065 & 173.5$\pm$0.2 & 1.857$\pm$0.044 & & \\
2017-06-17 & 15.4 & C & 373.7$\pm$0.5 & 0 & 0 & 0.062$\pm$0.000 & 6.79 & 13.58 \\
 & & J4 & 144.0$\pm$0.37 & 0.484$\pm$0.001 & 178.5$\pm$0.0 & 0.311$\pm$0.000 & & \\
 & & J3 & 8.9$\pm$0.47 & 1.157$\pm$0.014 & $-$141.2$\pm$0.6 & 0.212$\pm$0.011 & & \\
 & & J2 & 41.3$\pm$1.30 & 1.718$\pm$0.016 & $-$172.3$\pm$0.1 & 0.931$\pm$0.011 & & \\
 & & J1 & 20.9$\pm$3.65 & 3.941$\pm$0.099 & 173.8$\pm$0.3 & 1.815$\pm$0.065 & & \\
2018-04-22 & 15.4 & C & 646.6$\pm$0.8 & 0 & 0 & 0.092$\pm$0.000 & 5.28 & 10.56 \\
 & & J4 & 151.7$\pm$0.65 & 0.561$\pm$0.001 & 177.7$\pm$0.0 & 0.301$\pm$0.001 & & \\
 & & J3 & 13.5$\pm$1.03 & 1.263$\pm$0.022 & $-$143.3$\pm$0.8 & 0.204$\pm$0.017 & & \\
 & & J2 & 63.7$\pm$2.44 & 1.726$\pm$0.021 & $-$174.5$\pm$0.1 & 1.019$\pm$0.014 & & \\
 & & J1 & 18.9$\pm$3.66 & 4.134$\pm$0.112 & 173.6$\pm$0.3 & 1.669$\pm$0.074 & & \\
2018-05-31 & 15.4 & C & 744.8$\pm$1.1 & 0 & 0 & 0.097$\pm$0.000 & 5.50 & 11.01 \\
 & & J4 & 136.7$\pm$0.87 & 0.579$\pm$0.001 & 178.9$\pm$0.0 & 0.287$\pm$0.001 & & \\
 & & J2 & 64.0$\pm$2.11 & 1.717$\pm$0.017 & $-$174.4$\pm$0.1 & 1.037$\pm$0.011 & & \\
 & & J1 & 19.0$\pm$4.13 & 4.154$\pm$0.124 & 172.9$\pm$0.3 & 1.721$\pm$0.082 & & \\
2018-07-21 & 8.7 & C & 632.8$\pm$1.2 & 0 & 0 & 0.179$\pm$0.001 & 4.29 & 8.57 \\
2018-12-16 & 15.4 & C & 791.3$\pm$2.1 & 0 & 0 & 0.110$\pm$0.001 & 4.50 & 9.01 \\
 & & J4 & 144.1$\pm$1.54 & 0.592$\pm$0.002 & 177.5$\pm$0.0 & 0.317$\pm$0.001 & & \\
 & & J2 & 53.5$\pm$2.78 & 1.886$\pm$0.027 & $-$173.2$\pm$0.1 & 0.853$\pm$0.018 & & \\
 & & J1 & 22.0$\pm$9.14 & 3.925$\pm$0.245 & 172.2$\pm$0.7 & 2.352$\pm$0.169 & & \\
\enddata                  
\tablecomments{
(1) Observation date; 
(2) Frequency; 
(3) Component label; 
(4) Flux density; 
(5--6) Radial distance and position angle of the jet component with respect to the core; 
(7) Size of fitted Gaussian component; 
(8) Brightness temperature of the core in $10^{11}$ Kelvin (K); 
(9) Doppler factor.
}
\end{deluxetable}

\begin{figure}
\centerline{\includegraphics[width=\linewidth]{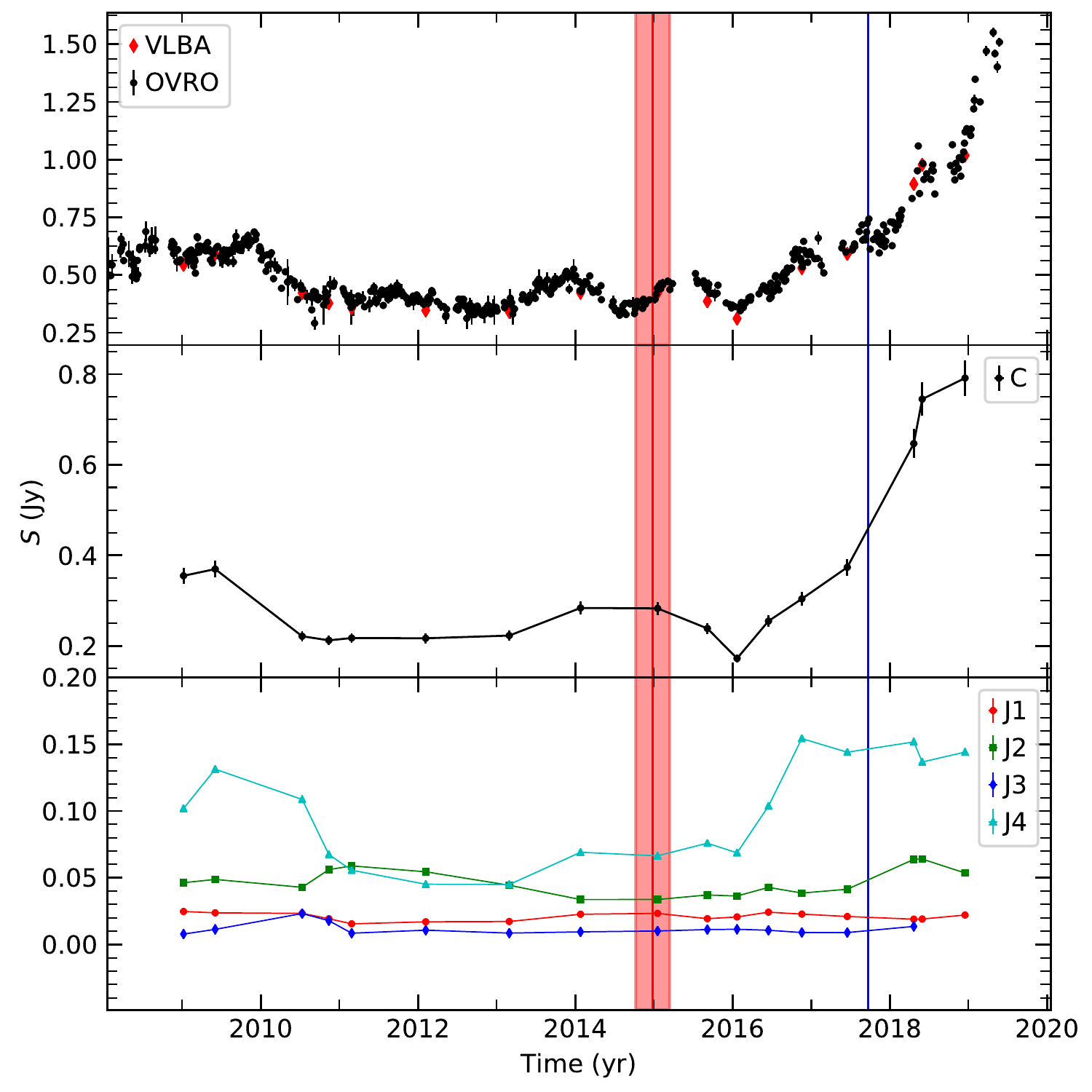}}
\caption{
Radio light curve of \txs{} observed with the Owens Valley Radio Observatory (OVRO) 40~m telescope ({\it top panel}). The integrated flux densities derived from the 15-GHz VLBA data are overlaid and shown as {\it red diamond} symbol. The blue-colored vertical line marks the epoch of the discovery of neutrino--$\gamma$-ray flare on 22 September 2017 \citep{2018Sci...361.1378I}. The red-colored shaded vertical bar indicates the time span between September 2014 and March 2015 in which a $3.5\sigma$ evidence was found at the position of \txs{} from historical IceCube data \citep{2018Sci...361..147I}. The {\it middle panel} shows the flux density variation of the core with time, and the {\it bottom panel} shows the flux density variations of the jet components with time. The core and jet flux densities are referred to Table~\ref{tab:comp}.
}
\label{fig5}
\end{figure}

\end{document}